\documentstyle[11pt]{article}
\begin{document}
\begin{center}
{\LARGE {\bf Dashen's theorem and electromagnetic masses of the mesons}}\\[5mm]
Dao-Neng Gao and Mu-Lin Yan\\
{\small Chinese Center for Advanced Science and Technology(World Lab)\\
P.O.Box 8730, Beijing,100080,P.R.China}\\
{\small Center for Fundamental Physics,
University of Science and Technology of China\\
Hefei, Anhui 230026, P.R.China}\footnote{Mailing address}\\[2mm]
Bing-An Li\\
{\small Department of Physics and Astronomy, University of Kentucky\\
Lexington, Kentucky 40506 USA}\\
\end{center}
\vspace{8mm}
\begin{abstract}
\noindent
Employing $U(3)_L\times U(3)_R$ chiral field theory, we find that Dashen's
theorem, which holds for pseudoscalar mesons, can be generalized to the
sector of axial-vector mesons, however, fails for vector mesons.
\end{abstract}

\vspace{1mm}
\noindent
{\bf PACS} number(s): 13.40.Dk,11.30.Rd,12.40.Vv,14.40.--n

\noindent
keywords:$\;\;$ chiral symmetry, Dashen's theorem, electromagnetic masses,
           vector meson dominance.
\vspace{3mm}

1. Chiral symmetry plays an important role in describing the low energy
hadronic physics. Based on chiral symmetry in three-flavor space, there
are three multiplets due to spontaneous symmetry breaking: an octet
of massless pseudoscalar mesons $0^{-}$($\pi $, $K$, $\eta $), and two
massive multiplets with quantum number $1^{-}$($\rho $, $\omega $, $K^{*}$
and $\phi$) and $1^{+}$($a_1$, $K_1$, $f$ and $f_s$). Significant SU(3) symmetry
breaking will result in the mass splittings of these low-lying mesons in the
same multiplet. Also, the energy of the virtual photon cloud around these
mesons can cause the mass differences between charged particles and their
corresponding neutral partners.

The earliest study on electromagnetic masses of mesons has been done by
Dashen\cite{Da} nearly thirty years before. Dashen's theorem, which states that the
square electromagnetic mass difference between the charge pseudoscalar
mesons and their corresponding neutral partners are equal in the chiral
SU(3) limit, is expressed as 
\begin{eqnarray}
& &(m_{\pi^\pm}^2-m_{\pi^0}^2)_{EM}=(m_{K^\pm}^2-m_{K^0}^2)_{EM},\nonumber \\
& &(m_{\pi^0}^2)_{EM}=0,\;\;\;\;(m_{K^0}^2)_{EM}=0.
\end{eqnarray}
The subscript EM denotes electromagnetic mass.

The three multiplets($0^-$, $1^-$ and $1^+$)
correspond to the same quark constituent but different spin or
parity in the quark model. The mass gaps between them are due to spontaneous
chiral symmetry breaking. Thus a very natural question is to ask whether
Dashen's theorem(which holds for pseudoscalar mesons) can be generalized to
be obeyed by vector and axial-vector mesons. In this paper, by employing
$U(3)_L\times U(3)_R$ chiral theory\cite{Li1,Li2}, it will be shown that the
generalization to axial-vector mesons is valid, or, in the chiral SU(3)
limit, 
\begin{eqnarray}
& &(m_{a^\pm}^2-m_{a^0}^2)_{EM}=(m_{K_1^\pm}^2-m_{K_1^0}^2)_{EM},\nonumber \\
& &(m_{a^0}^2)_{EM}=0,\;\;\;\;(m_{K_1^0}^2)_{EM}=0.
\end{eqnarray}
However, similar generalization fails for vector mesons.
The latter is the same conclusion as one given by
Bijnens and Gosdzinsky\cite{JG} based on
the heavy vector meson chiral Lagrangian\cite{JMW,EBH,Ge}, but Eq.(2) is new.

The formalism of $U(3)_L\times U(3)_R$ chiral field theory of pseudoscalar,
vector, and axial-vector mesons has been described in Refs.\cite{Li1,Li2}.
This theory is of self-consistency and phenomenological success, and
provides a unified description
of meson physics at low energies.
Because various meson-processes or properties related to strong,
electromagnetic and weak interactions have been calculated in
Refs.\cite{Li1,Li2} and the results
are in good agreement with the experimental data, all chiral coefficients
$L_1, ..., L_{10}$ in the chiral perturbation theory($\chi$PT)\cite{SGD,GL}
corresponding to these processes can be well determined.
Actually, two of them ($\alpha_1=4L_1+2L_3, \alpha_2=4L_2$) have already
been shown in\cite{Li1}. Therefore the present theory is believed to
be consistent with both $\chi$PT and the chiral coupling theory
of resonances\cite{EGPR,EGLPR}. A systematical study on this issue will
be presented in detail elsewhere.
In the present paper, we focus our attention on the investigation upon  the
electeomagnetic masses of the mesons and the generalization of Dashen's theorem
in the framework of $U(3)_L\times U(3)_R$ chiral theory.

In the $U(3)_L\times U(3)_R$ chiral theory,
vector meson dominance(VMD)\cite{JS} in the meson physics is a
natural result of this theory instead of an input. Namely, the dynamics of
the electromagnetic interactions of mesons has been introduced and
established naturally. Therefore, the present theory makes it possible to
evaluate the electromagnetic self-energies of these low-lying mesons
systematically. According to this pattern, for example, we can work out 
the well known old result of $(m_{\pi^\pm}^2-m_{\pi^0}^2)_{EM}$ given 
by Das et al\cite{DGMLY}, which serves as the leading order in our more
accurate evaluations (to see below). This indicates that our pattern
of evaluating the electromagnetic self-energies of mesons is consistent with the
known theories in the $\pi-$meson case under the lowest energies limit.
Since the dynamics of mesons, including pseudoscalar, vector and 
axial-vector, is described in the present theory, the method of calculating
electromagnetic masses in this paper is legitimate not only for $\pi$ and $K-$mesons,
but also for $a_1$, $K_1$, $\rho$ and $K^*-$mesons. Thus, both
checking the Dashen's argument of Eq.(1) in the framework of the effective
quantum fields theory and investigating its generalizations mentioned
above become practical.

The basic Lagrangian of this chiral field theory is (hereafter we use the
notations in Refs.\cite{Li1,Li2})
\begin{eqnarray}
\lefteqn{{\cal L}=\bar{\psi}(x)(i\gamma\cdot\partial+\gamma\cdot v
+\gamma\cdot a\gamma_{5}
-mu(x))\psi(x)}\nonumber \\
& &+{1\over 2}m^{2}_{1}(\rho^{\mu}_{i}\rho_{\mu i}+
\omega^{\mu}\omega_{\mu}+a^{\mu}_{i}a_{\mu i}+f^{\mu}f_{\mu})\nonumber\\
& &+{1\over2}m^2_2(K_{\mu}^{*a}K^{*a\mu}+K_1^{\mu}K_{1\mu})\nonumber \\
& &+\frac{1}{2}m^2_3(\phi_{\mu} \phi^{\mu}+f_s^{\mu}f_{s\mu})
\end{eqnarray}
where $u(x)=exp[i\gamma_{5} (\tau_{i}\pi_{i}+\lambda_a K^a+\eta
+\eta^{\prime})]$($i$=1,2,3 and $a$=4,5,6,7), $a_{\mu}=\tau_{i}a^{i}_{\mu}+
\lambda_a K^a_{1\mu}+(\frac{2}{3} +\frac{1}{\sqrt{3}} \lambda_8)f_{\mu}+(
\frac{1}{3}-\frac{1}{\sqrt{3}} \lambda_8)f_{s\mu},
v_{\mu}=\tau_{i}\rho^{i}_{\mu}+\lambda_a K_{\mu}^{*a}+(\frac{2}{3}+ \frac{1}{
\sqrt{3}}\lambda_8)\omega_{\mu}+(\frac{1}{3}- \frac{1}{\sqrt{3}}%
\lambda_8)\phi_{\mu}$. The $\psi$ in Eq.(1) is $u$,$d$,$s$ quark fields. $m$
is a parameter related to the quark condensate. Here, the mesons are bound
states in QCD, and they are not fundamental fields. Therefore, in Eq.(1) there
are no kinetic terms for these fields and the kinetic terms will be
generated from quark loops.

Following Refs.\cite{Li1,Li2}, the normal part and abnormal part of the effective
Lagrangians, i.e. ${\cal L}_{RE}$ and ${\cal L}_{IM}$, can be obtained by
performing path integration over quark fields. The dynamics of the mesons
can be read off from ${\cal L}_{RE}$ and ${\cal L}_{IM}$. It is natural
to obtain VMD in the present theory, which reads\cite{Li2}
\begin{eqnarray}
{\cal L}_{\rho \gamma}=-{e \over f_\rho} \partial_\mu
    \rho_\nu^0 (\partial^\mu A^\nu -\partial^\nu A^\mu),\nonumber \\
{\cal L}_{\omega \gamma}=-{e \over f_\omega} \partial_\mu
    \omega_\nu (\partial^\mu A^\nu -\partial^\nu A^\mu),\nonumber \\
{\cal L}_{\phi \gamma}=-{e\over f_\phi}  \partial_\mu
    \phi_\nu (\partial^\mu A^\nu-\partial^\nu A^\mu).
\end{eqnarray}
The direct couplings of photon to other mesons are constructed through the
substitutions 
\begin{eqnarray}
\rho_\mu^0\rightarrow \frac{e}{f_\rho} A_\mu,\;\;\;
\omega_\mu\rightarrow \frac{e}{f_\omega} A_\mu,\;\;\;
\phi_\mu\rightarrow \frac{e}{f_\phi} A_\mu.
\end{eqnarray}
where 
\begin{eqnarray}
\frac{1}{f_\rho}=\frac{1}{2}g,\;\;\;\frac{1}{f_\omega}=\frac{1}{6}g,
\;\;\;\frac{1}{f_\phi}=-\frac{1}{3\sqrt{2}}g.
\end{eqnarray}
$g$ is a universal coupling constant in this theory. It can be determined by
taking the experimental value of $m_a$ as an input\cite{Li1,Li2,GYL}.

Using ${\cal L}_i (\Phi,\gamma,...)|_{\Phi=\pi, a, v,...}$(which can be
obtained from ${\cal L}_{RE}$ or ${\cal L}_{IM}$), we can evaluate the following
S-matrix 
\begin{equation}
S_{\Phi}=\langle \Phi |T{\rm exp}[i\int dx^4 {\cal L}_i
(\Phi,\gamma,...)]-1|\Phi\rangle |_{\Phi=\pi, a, v,...}. 
\end{equation}
On the other hand $S_\Phi$ can also be expressed in terms of the effective
Lagrangian of $\Phi$ as 
$$
S_\Phi=\langle \Phi |i\int d^4 x {\cal L}_{{\rm eff}} (\Phi) | \Phi \rangle. 
$$
Noting ${\cal L}={\frac{1}{2}}\partial_\mu\Phi\partial^\mu \Phi-{\frac{1}{2}}
m_{\Phi}^2 \Phi^2$, then the electromagnetic interaction correction to the
mass of $\Phi$ reads 
\begin{equation}
\delta m_\Phi^2 ={\frac{2iS_\Phi }{\langle \Phi |\Phi^2 | \Phi \rangle }}. 
\end{equation}
where $\langle \Phi |\Phi^2 |\Phi \rangle = \langle \Phi |\int d^4 x
\Phi^2(x) |\Phi \rangle $. Thus, all of virtual photon contributions to mass
of the mesons can be calculated systematically.

In the following, firstly, we shall re-examine Dashen's theorem for
pseudoscalar mesons in the framework of the present theory. Then
the generalization of this theorem for axial-vector and vector mesons
is studied. Finally, we give the summary and conclusions.

\vspace{8mm}
2. Due to VMD(Eq.(4)), the interaction Lagrangians which
contribute to electromagnetic mass of the mesons have to contain the neutral
vector meson fields which only include $\rho^0$, $\omega$ and $\phi$. When
the calculations are of $O(\alpha_{em})$ and one-loop, there are two sorts
of vertices contributing to electromagnetic self-energies of pseudoscalar
mesons: four points vertices and three points vertices. The former must
be the coupling of two pseudoscalar fields and two neutral vector mesons
fields, and the latter must be the interaction of a pseudoscalar field to a
neutral vector meson plus another field.

Thus, from Refs.\cite{Li1,Li2}, the interaction Lagrangians contributing to
electromagnetic mass of pseudoscalar mesons($\pi$ and $K$) in the chiral
limit can be easily obtained, which reads as follows 
\begin{eqnarray}
& &{\cal L}_{\rho \rho \pi \pi}={4F^2\over g^2 f_\pi^2}(\rho_\mu^0
\rho^{0\mu}+{1 \over 2\pi^2 F^2 }\partial_\nu\rho_\mu^0 \partial^\nu\rho^{0\mu})
  \pi^+\pi^-,\\
& &{\cal L}_{\rho \pi a}={2i\gamma F^2 \over f_\pi g^2}\rho_\mu^0
\pi^+ (a^{-\mu}-{1 \over 2\pi^2 F^2}\partial^2 a^{-\mu})+h.c.,  \\
& &{\cal L}_{K^+K^-v v}=\frac{F^2}{f_\pi^2 g^2}\{
 (\rho^{0\mu}+v_\mu^8)^2+\frac{1}{2\pi^2 F^2}(\partial_\nu\rho^{0\mu}
 +\partial^\nu v_\mu^8)^2\}K^+K^{-},\\
& &{\cal L}_{K^{\pm} K_1^{\pm} v}=\frac{i\gamma F^2}{g^2 f_\pi}(\rho_\mu^0+v_\mu^8)
K^+ (K_1^{-\mu}-\frac{1}{2\pi^2 F^2}\partial^2 K_1^{-\mu})+h.c.,\\
& &{\cal L}_{K^0\bar{K}^0 v v}=\frac{F^2}{f_\pi^2 g^2}\{
 (-\rho^{0\mu}+v_\mu^8)^2+\frac{1}{2\pi^2 F^2}(-\partial_\nu\rho^{0\mu}
 +\partial^\nu v_\mu^8)^2\}K^0\bar{K}^0,\\
& &{\cal L}_{K^{0} K_1^{0} v}=\frac{i\gamma F^2}{g^2 f_\pi}(-\rho_\mu^0+v_\mu^8)
K^0 (\bar{K}_1^{0\mu}-\frac{1}{2\pi^2 F^2}\partial^2 \bar{K}_1^{0\mu})+h.c.
\end{eqnarray}
where 
\begin{eqnarray*}
\pi^\pm=\frac{1}{\sqrt{2}}(\pi^1\pm i\pi^2),\;\;\; a^\pm=\frac{1}{\sqrt{2}}
(a^1\pm i a^2),\\
K^\pm=\frac{1}{\sqrt{2}}(K^4\pm i K^5),\;\;\; K^0(\bar{K}^0)=
\frac{1}{\sqrt{2}}(K^6\pm i K^7),\\
K_{1\mu}^\pm=\frac{1}{\sqrt{2}}(K^4_{1\mu} \pm i K^5_{1\mu}),\;\;\;
K_{1\mu}^0(\bar{K}_{1\mu}^0)=\frac{1}{\sqrt{2}}(K^6_{1\mu}\pm i
K^7_{1\mu}).
\end{eqnarray*}
with 
\begin{eqnarray}
& &F^2={f_\pi^2 \over 1-\frac{2c}{g}},\;\;\;c={f_\pi^2\over 2 g m_\rho^2},\\
\nonumber
& &\gamma=(1-\frac{1}{2\pi^2 g^2})^{-1/2}.
\end{eqnarray}
where $v$ denotes the neutral vector mesons $\rho^0$, $\omega$ and $\phi$,
$v_\mu^8=\omega_\mu-\sqrt{2}\phi_\mu$. $
f_\pi$ is pion's decay constant, and $f_\pi=0.186GeV$.
Here, we neglect the contributions to
electromagnetic mass of pions or kaons which are proportional to $m_\pi^2$
or $m_K^2$, because we are only interested in the case of chiral limit.

Note that there are no contributions to electromagnetic masses of $\pi^0$ in the
chiral limit. This means 
\begin{equation}
({m_{\pi^0}^2})_{EM}=0,\;\;\;\;{for\;\; massless\;\; quark.} 
\end{equation}

The major difference of the kaon and pion Lagrangians is that in the
kaon case all the three nonet vector resonances $\rho^0,\omega$ and $\phi$
contribute, which results in a contribution to the neutral kaon electromagnetic
self-energy. This contribution vanishes in the 
SU(3)-vector-meson symmetry limit that the three vector
masses are equal, i.e. $m_\rho=m_\omega=m_\phi$(to see below). These results
are in agreement with ones given by Donoghue and Perez in Ref.\cite{DP}
in the framework of $\chi$PT.

The Lagrangians contributing to electromagnetic masses of $K^\pm$ are
different from ones to electromagnetic mass of $K^0$. The difference comes
from the structure constants of SU(3) group: $f_{345}=-f_{367}=\frac{1}{2},
f_{458}=f_{678}=\frac{\sqrt{3}}{2}$. Note that the vector meson fields( $\rho
$, $\omega$ and $\phi$) and axial-vector meson fields($a_1$ and $K_1(1400)$)
in the above Lagrangians would always appear as propagators in the
S-matrices which can contribute the electromagnetic self-energies of
pseudoscalar mesons. Using Eq.(4) with SU(3) symmetry limit $
m_\rho=m_\omega=m_\phi$, we can easily obtain that $v_\mu^8$ is equivalent
to $\rho_\mu^0$ in calculating the electromagnetic masses of the mesons.
Thus, the Lagrangians contributing to electromagnetic masses of $K^0$ will
vanish, namely, 
\begin{equation}
(m_{K^0}^2)_{EM}=0,\;\;\;\;{in\;\; the\;\; chiral\;\; SU(3)\;\; limit.} 
\end{equation}
Likewise, we can conclude that the Lagrangian (11) and (12) are entirely
equivalent to Lagrangians (9) and (10) respectively under the limit of $
m_\rho=m_\omega=m_\phi$ and $m_a=m_{K_1}$ . Then, according to Eqs.(7) and
(8) we have
\begin{equation}
(m_{K^\pm}^2)_{EM}=(m_{\pi^\pm}^2)_{EM}, \;\;\;\; {in\;\;the\;\;chiral\;\;SU(3)
\;\;limit.} 
\end{equation}

Above arguments and conclusions can also be checked by practical 
calculations which are standard in quantum fields theory.
From Eqs.(9)-(14) and VMD, the electromagnetic masses of $\pi^\pm$, $K^\pm$ and
$K^0$ can be derived(to see Ref.\cite{GYL} for details), which are as follows,
\begin{equation}
(m_{\pi^\pm}^2)_{EM} =i{\frac{e^2 }{f_\pi^2}}\int\frac{d^4 k}{(2\pi)^4}
(D-1)m_\rho^4 {\frac{{(F^2+{\frac{k^2 }{2\pi^2}})} }{{k^2 (k^2-m_\rho^2)^2 }}
} [1+{\frac{\gamma^2 }{g^2}}{\frac{{F^2+{\frac{k^2 }{2\pi^2}}} }{{k^2 -m_a^2}
}}] 
\end{equation}
\begin{eqnarray}
(m_{K^\pm}^2)_{EM}-(m_{K^0}^2)_{EM}&=&i\frac{e^2}{f_\pi^2}\int\frac{d^4 k}{(2\pi)^4}
(D-1)(F^2+\frac{k^2}{2\pi^2})(1+\frac{\gamma^2}{g^2}\frac{F^2
+\frac{k^2}{2\pi^2}}{k^2-m_{K_1}^2})\nonumber \\
& &\times[\frac{1}{3}\frac{m_\rho^2 m_\omega^2}
{k^2(k^2-m_\rho^2)(k^2-m_\omega^2)}+\frac{2}{3}\frac{m_\rho^2 m_\phi^2}
{k^2(k^2-m_\rho^2)(k^2-m_\phi^2)}]
\end{eqnarray}
\begin{eqnarray}
(m_{K^0}^2)_{EM}&=&i\frac{e^2}{4 f_\pi^2}\int\frac{d^4 k}{(2\pi)^4}
(D-1)(F^2+\frac{k^2}{2\pi^2})(1+\frac{\gamma^2}{g^2}\frac{F^2
+\frac{k^2}{2\pi^2}}{k^2-m_{K_1}^2})k^2\nonumber\\
& &\times[\frac{1}{k^2-m_\rho^2}-\frac{1}{3}\frac{1}{k^2-m_\omega^2}-
\frac{2}{3}\frac{1}{k^2-m_\phi^2}]^2
\end{eqnarray}
where $D=4-\varepsilon$.
Obviously, taking $m_\rho=m_\omega=m_\phi$, and $m_a=m_{K_1}$, Eq.(20) is
exactly Eq.(19). Also, the contribution of Eq.(21) is
zero with $m_\rho=m_\omega=m_\phi$. Thus, Eq.(18) holds, and Dashen's
theorem(Eq.(1)) is automatically obeyed in the chiral SU(3) limit.

In the above, the calculations on electromagnetic masses are up to $O(p^4)$ (or up to
4-order derivative operators in effective Lagrangians). In the remainder
of this Section, we would like to back to $O(p^2)$ in order to show the
leading order of electromagnetic mass difference of $\pi-$meson,
$(m_{\pi^\pm}^2-m_{\pi^0}^2)_{EM}^{(0)}$, and to see if it is
just the old result of Das et al\cite{DGMLY} or not. To $O(p^2)$,
the interaction Lagrangians ${\cal L}_{\rho\rho\pi\pi}$ and
${\cal L}_{\rho\pi a}$(Eqs.(9)(10)) will be simplified as follows
\begin{eqnarray*}
& &{\cal L}_{\rho\rho\pi\pi}=\frac{4F^2}{g^2 f_\pi^2}\rho^0_\mu \rho^{0\mu}
\pi^+\pi^-,\;\;\;\;\;
{\cal L}_{\rho\pi a}=\frac{2iF^2}{f_\pi g^2}\rho^0_{\mu}\pi^+ a^{-\mu}+h.c..
\end{eqnarray*}
Thus, in the chiral limit, the electromagnetic self-energy of pions is
\begin{equation}
(m_{\pi^\pm}^2-m_{\pi^0}^2)_{EM}^{(0)}=(m_{\pi^\pm}^2)_{EM}^{(0)}
=i\frac{3e^2}{f_\pi^2}
\int\frac{d^4k}{(2\pi)^4}m_\rho^4\frac{F^2}{k^2(k^2-m_\rho^2)^2}
(1+\frac{F^2}{g^2(k^2-m_a^2)})
\end{equation}
The Feynman integration in Eq.(22) is finite. So it is
straightforward to get the result of $(m_{\pi^\pm}^2-m_{\pi^0}^2)_{EM}$
after performing this integration, which is
\begin{equation}
(m_{\pi^\pm}^2-m_{\pi^0}^2)_{EM}^{(0)}
=\frac{3\alpha_{em}m_\rho^4}{8\pi f_\pi^2}
\{\frac{2F^2}{m_\rho^2}-\frac{2F^4}{g^2(m_a^2-m_\rho^2)}(\frac{1}{m_\rho^2}
+\frac{1}{m_a^2-m_\rho^2}log{\frac{m_\rho^2}{m_a^2}})\}
\end{equation}
where $\alpha_{em}=\frac{e^2}{4\pi}$. Because we only consider the
second order derivative terms in the
Largrangian, the relation between $m_a$ and $m_\rho$ is $m_a^2=\frac{F^2}{g^2}
+m_\rho^2$ instead of Eq.(27) in Ref.\cite{Li1}. Combining this relation
with Eq.(15), Eq.(23) can be simplified
\begin{equation}
(m_{\pi^\pm}^2-m_{\pi^0}^2)_{EM}^{(0)}
=\frac{3\alpha_{em}}{4\pi}\frac{m_a^2 m_\rho^2}{m_a^2-m_\rho^2}
log{\frac{m_a^2}{m_\rho^2}}
\end{equation}
It has been pointed out that
Kawarabayashi-Suzuki-Riazuddin-Fayyazuddin(KSRF) sum rule\cite{KSRF} is
satisfied reasonably well in the present theory\cite{Li1}. Thus, using
Eq.(53)($2f_\pi^2=g^2 m_\rho^2$) in Ref.\cite{Li1}, we can get
$$
m_a^2=2 m_\rho^2
$$
This relation is consistent with the sencond Weinberg sum
rule\cite{Wb}. Now, Eq.(24) is
\begin{equation}
(m_{\pi^\pm}^2-m_{\pi^0}^2)_{EM}^{(0)}
=\frac{3log2}{2\pi}\alpha_{em} m_\rho^2
\end{equation}
which is exactly the result given by Das et al\cite{DGMLY}, and serves
as the leading term of Eq.(19) in the lowest energies limit.

\vspace{8mm}
3. It is straightforward to extend the present study to the
case of axial-vector mesons. In the same way as in the case of pions and
kaons, the Lagrangians contributing to electromagnetic masses of
axial-vector mesons($a_1$ and $K_1$) read 
\begin{eqnarray}
{\cal L}_{a a \rho \rho}&=&-\frac{4}{g^2}[
\rho_\mu^0 \rho^{\mu 0} a_\nu^+ a^{-\nu}-\frac{\gamma^2}{2}\rho_\mu^0
\rho_\nu^0(a^{+\mu}a^{-\nu}+a^{-\mu}a^{+\nu})],\\
{\cal L}_{a a \rho}&=&\frac{2i}{g}(1-\frac{\gamma^2}{\pi^2 g^2})
\partial^\nu \rho_\mu^0 a^{+\mu}a_{\nu}^--\frac{2i}{g}\rho_\nu^0 a^{+\mu}
(\partial^\nu a_{\mu}^--\gamma^2 \partial_\mu a^{-\nu})+h.c.,\\
{\cal L}_{a \pi \rho}&=&\frac{2i}{g}(\beta_1 \rho_\mu^0 \pi^+ a^{-\mu}
+\beta_2 \rho_\nu^0 \partial^{\mu\nu}\pi^+ a_{\mu}^-
+\beta_3 \rho_\mu^0 a^{+\mu}\partial^2 \pi^--\beta_4 \rho_\mu^0
\pi^+\partial^2 a^{-\mu}\nonumber \\
& &-\beta_5\rho_\mu^0\partial_\nu a_{\mu}^+ \partial^\nu \pi^-)+h.c.
\end{eqnarray}
\begin{eqnarray}
{\cal L}_{K_1^+ K_1^- v v}&=&-\frac{1}{g^2}[(\rho_\mu^0+v^{8\mu})^2
K_{1\nu}^+ K_1^{-\nu} \nonumber\\
& &-\frac{\gamma^2}{2}(\rho_\mu^0+v_{\mu}^8)(\rho_\nu^0+v_\nu^8)
(K_1^{+\mu} K_1^{-\nu}+K_1^{-\mu}K_1^{+\nu})], \\
{\cal L}_{K_1^+ K_1^- v}&=&\frac{i}{g}(1-\frac{\gamma^2}{\pi^2 g^2})
(\partial^\nu\rho_{\mu}^0+\partial^\nu v_\mu^8)K_1^{+\mu}K_{1\nu}^{-}
\nonumber \\
& &-\frac{i}{g}(\rho_\nu^0+v_\nu^8)[K_1^{+\mu}(\partial^\nu K_{1\mu}^--\gamma^2
\partial_{\mu}K_{1}^{-\nu})
+ h.c.,  \\
{\cal L}_{K^\pm K_1^\pm v}&=&\frac{i}{g}\beta_1
(\rho_\mu^0+v_\mu^8)K^{+}K_1^{-\mu}
+\frac{i}{g}\beta_2(\rho_\nu^0+v_\nu^8)\partial^{\mu\nu} K^+K_{1\mu}^-\nonumber\\
& &+\frac{i}{g}\beta_3(\rho_\mu^0+v_\mu^8)K_1^{+\mu}\partial^2 K^{-}
-\frac{i}{g}\beta_4(\rho_\mu^0+v_\mu^8)K^{+}\partial^2 K_1^{-\mu}\nonumber \\
& &-\frac{i}{g}\beta_5(\rho_\mu^0+v_\mu^8)\partial_\nu K_{1\mu}^+ \partial^\nu K^-
+h.c.
\end{eqnarray}
\begin{eqnarray}
& &{\cal L}_{K_1^0 \bar{K}_1^0 v v}={\cal L}_{K_1^+ K_1^- v v}\{
\rho^0\leftrightarrow -\rho^0,K_1^\pm\leftrightarrow K_1^0(\bar{K}_1^0)\},
\\
& &{\cal L}_{K_1^0 \bar{K}_1^0 v }={\cal L}_{K_1^+ K_1^-  v}\{
\rho^0\leftrightarrow -\rho^0,K_1^\pm\leftrightarrow K_1^0(\bar{K}_1^0)\},
\\
& &{\cal L}_{K_1^0 K^0 v }={\cal L}_{K^\pm K_1^\pm v }\{
\rho^0\leftrightarrow -\rho^0,K_1^\pm\leftrightarrow K_1^0(\bar{K}_1^0)
,K^\pm\leftrightarrow K^0(\bar{K}^0)\}.
\end{eqnarray}
with 
\begin{eqnarray*}
\beta_1=\frac{\gamma F^2}{g f_\pi},\;\;\;
\beta_2=\frac{\gamma}{2\pi^2 g f_\pi},\\
\beta_3=\frac{3\gamma}{2\pi^2 g f_\pi}(1-\frac{2c}{g})+\frac{2\gamma c}{f_\pi},
\;\;\;\beta_4=\frac{\gamma}{2\pi^2 g f_\pi},\\
\beta_5=\frac{3\gamma}{2\pi^2 g f_\pi}(1-\frac{2 c}{g})+\frac{4\gamma c}{f_\pi}.
\end{eqnarray*}
From the above Lagrangians(Eqs.(26)-(34)), we can find the case of axial-vector mesons is
very like that of pseudoscalar mesons. Therefore, similar analyses can yield
the similar results for axial-vector mesons, i.e. in the chiral SU(3) limit,
\begin{eqnarray}
& &(m_{a^\pm}^2-m_{a^0}^2)_{EM}=(m_{K_1^\pm}^2-m_{K_1^0}^2)_{EM},\nonumber \\
& &(m_{a^0}^2)_{EM}=0,\;\;\;\;(m_{K_1^0}^2)_{EM}=0.
\end{eqnarray}
Here, it is necessary to check Eq.(35) by explicit calculations which can be
done in the same way as in the previous section. Using Eqs.(26)-(34) together
with VMD, we can get the expressions for the electromagnetic masses of
axial-vector mesons, which are as follows,
\begin{equation}
(m_{a^0}^2)_{EM}=0
\end{equation}
$$
(m_{a^\pm}^2)_{EM}=(m_{a^\pm}^2)_{EM}(1)+(m_{a^\pm}^2)_{EM}(2)+(m_{a^\pm}^2)_{EM}(3)
$$
with
\begin{eqnarray}
&&(m_{a^\pm}^2)_{EM}(1)
=ie^2\frac{\gamma^2\langle a| \int d^4 x
a^{\underline{i}\mu}a^{\underline{i}\nu}|a\rangle-\langle a| \int d^4 x
a^{\underline{i}\lambda}a_{\lambda}^{\underline{i}}|a\rangle
g^{\mu\nu}}{\langle a| \int d^4 x a_{\mu}^{\underline{i}}a^{\underline{i}\mu}
|a\rangle}
\nonumber \\
& &\times\int\frac{d^4 k}{(2\pi)^4}\frac{m_\rho^4}{k^2(k^2-m_\rho^2)^2}
(g_{\mu\nu}-\frac{k_\mu k_\nu}{k^2}),\\
\mbox{\vspace{2mm}}
&&(m_{a^\pm}^2)_{EM}(2)
=\frac{ie^2}{\langle a| \int d^4 x a^{\underline{i}\mu}
a_{\mu}^{\underline{i}}|a\rangle}\int\frac{d^4 k}{(2\pi)^4}\frac{1}{k^2-2p\cdot k}
\frac{m_\rho^4}{k^2(k^2-m_\rho^2)^2}\nonumber \\
& &\times\{\langle a| \int d^4 x a^{\underline{i}\mu}a_{\mu}^{\underline{i}}|a\rangle
[4 m_a^2+(b^2+2b\gamma^2)k^2
+2\gamma^4 p\cdot k-\frac{4(p\cdot k)^2}{k^2}\nonumber \\
& &-\frac{1}{m_a^2}(b k^2-(b-\gamma^2)p\cdot k)^2]
+\langle a| \int d^4 x a_{\mu}^{\underline{i}}a_{\nu}^{\underline{i}}|a\rangle
k^\mu k^\nu [-(3b^2-4 b+4)\nonumber \\
& &+D(b+\gamma^2)^2+4\gamma^2
-6b\gamma^2-2\gamma^4-\frac{2\gamma^4 p\cdot k}{k^2}
+\frac{1}{m_a^2 k^2}(b k^2-2(1-\gamma^2)p\cdot k)^2]\},\\
\mbox{\vspace{2mm}}
&&(m_{a^\pm})_{EM}(3)
=\frac{-ie^2}{\langle a| \int d^4 x a_{\mu}^{\underline{i}}
a^{\underline{i}\mu}|a\rangle}\int\frac{d^4 k}{(2\pi)^4}\frac{1}{(p-k)^2}
\frac{m_\rho^4}{k^2(k^2-m_\rho^2)^2}\nonumber\\
& &\times\{\langle a| \int d^4 x a_{\mu}^{\underline{i}}a^{\underline{i}\mu}|a\rangle
(\beta_1^\prime-3 \beta_2 p\cdot k+
\beta_3 k^2)^2+\nonumber \\
& &\langle a| \int d^4 x a_{\mu}^{\underline{i}}a_{\nu}^{\underline{i}}|a\rangle k^\mu k^\nu
[\beta_2 m_{a}^2-\frac{(\beta_1^\prime-2 \beta_2 p\cdot k+\beta_3 k^2)^2}{k^2}
]\}.
\end{eqnarray}
where $\underline{i}$=1,2.
\vspace{2mm}
\begin{eqnarray*}
(m_{K_1^\pm}^2)_{EM}-(m_{K_1^0}^2)_{EM}&=&[(m_{K_1^\pm}^2)_{EM}(1)-
(m_{K_1^0}^2)_{EM}(1)]+[(m_{K_1^\pm}^2)_{EM}(2)-(m_{K_1^0}^2)_{EM})(2)]\\
& &+[(m_{K_1^\pm}^2)_{EM}(3)-(m_{K_1^0}^2)_{EM}(3)]
\end{eqnarray*}
with
\begin{eqnarray}
&&(m_{K_1^\pm}^2)_{EM}(1)-(m_{K_1^0}^2)_{EM}(1)
=ie^2\frac{\gamma^2\langle K_1| \int d^4 x
K_1^{\mu+}K_1^{\nu-}|K_1\rangle-\langle K_1| \int d^4 x K_1^{\lambda+} K_{1\lambda}^-|K_1\rangle
g^{\mu\nu}}{\langle K_1| \int d^4 x K_{1\mu}^+K_1^{\mu-}|K_1\rangle}\nonumber \\
& &\times\int\frac{d^4 k}{(2\pi)^4}
(g_{\mu\nu}-\frac{k_\mu k_\nu}{k^2})
[\frac{1}{3}\frac{m_\rho^2 m_\omega^2}{k^2(k^2-m_\rho^2)(k^2-m_\omega^2)}+
\frac{2}{3}\frac{m_\rho^2 m_\phi^2}{k^2(k^2-m_\rho^2)(k^2-m_\phi^2)}],\\
\vspace{2mm}
&&(m_{K_1^\pm}^2)_{EM}(2)-(m_{K_1^0}^2)_{EM}(2)
=\frac{ie^2}{\langle K_1| \int d^4 x K_1^{\mu+}
K_{1\mu}^-|K_1\rangle}\int\frac{d^4 k}{(2\pi)^4}\frac{1}{k^2-2p\cdot k}\nonumber \\
& &\times\{\langle K_1| \int d^4 x K_1^{\mu+}K_{1\mu}^-|K_1\rangle[4 m_{K_1}^2+(b^2+2b\gamma^2)k^2
+2\gamma^4 p\cdot k-\frac{4(p\cdot k)^2}{k^2}\nonumber \\
& &-\frac{1}{m_{K_1}^2}(b k^2-(b-\gamma^2)p\cdot k)^2]
+\langle K_1| \int d^4 x K_{1\mu}^+K_{1\nu}^-|K_1\rangle k^\mu k^\nu [
-(3b^2-4 b+4)\nonumber \\
& &+D(b+\gamma^2)^2+4\gamma^2
-6b\gamma^2-2\gamma^4-\frac{2\gamma^4 p\cdot k}{k^2}
+\frac{1}{m_{K_1}^2 k^2}(b k^2-2(1-\gamma^2)p\cdot k)^2]\}\nonumber \\
& &\times[\frac{1}{3}\frac{m_\rho^2 m_\omega^2}{k^2(k^2-m_\rho^2)(k^2-
m_\omega^2)}+\frac{2}{3}\frac{m_\rho^2 m_\phi^2}{k^2(k^2-m_\rho^2)(
k^2-m_\phi^2)}],\\
\vspace{2mm}
&&(m_{K_1^\pm}^2)_{EM}(3)-(m_{K_1^0}^2)_{EM}(3)
=\frac{-ie^2}{\langle K_1| \int d^4 x K_{1\mu}^+
K_1^{\mu-}|K_1\rangle}\int\frac{d^4 k}{(2\pi)^4}\frac{1}{(p-k)^2-m_K^2}\nonumber\\
& &\times\{\langle K_1| \int d^4 x K_{1\mu}^+K_1^{\mu-}|K_1\rangle(\beta_1^\prime-3\beta_2 p\cdot k+
\beta_3 k^2)^2+\nonumber \\
& &\langle K_1| \int d^4 x K_{1\mu}^+K_{1\nu}^-|K_1\rangle k^\mu k^\nu
[\beta_2 m_{K_1}^2-\frac{(\beta_1^\prime-2 \beta_2 p\cdot k+\beta_3 k^2)^2}{k^2}
]\}\nonumber \\
& &\times[\frac{1}{3}\frac{m_\rho^2 m_\omega^2}{k^2(k^2-m_\rho^2)(k^2-
m_\omega^2)}+\frac{2}{3}\frac{m_\rho^2 m_\phi^2}{k^2(k^2-m_\rho^2)(
k^2-m_\phi^2)}].
\end{eqnarray}
\vspace{2mm}
$$
(m_{K_1^0}^2)_{EM}=(m_{K_1^0}^2)_{EM}(1)+(m_{K_1^0}^2)_{EM}(2)+(m_{K_1^0}^2)_{EM}(3)
$$
with
\begin{eqnarray}
&&(m_{K_1^0}^2)_{EM}(1)
=ie^2\frac{\gamma^2\langle K_1| \int d^4 x
K_1^{\mu 0}\bar{K}_1^{\nu 0}|K_1\rangle-\langle K_1| \int d^4 x K_1^{\lambda 0}
\bar{K}_{1\lambda}^0|K_1\rangle
g^{\mu\nu}}{4\langle K_1| \int d^4 x K_{1\mu}^0\bar{K}_1^{\mu 0}|K_1\rangle}\nonumber \\
& &\times\int\frac{d^4 k}{(2\pi)^4}
(k^2 g_{\mu\nu}-{k_\mu k_\nu})
[\frac{1}{k^2-m_\rho^2}-\frac{1}{3}\frac{1}{k^2-m_\omega^2}-
\frac{2}{3}\frac{1}{k^2-m_\phi^2}]^2,\\
&&(m_{K_1^0}^2)_{EM}(2)
=\frac{ie^2}{4\langle K_1| \int d^4 x K_1^{\mu 0}
\bar{K}_{1\mu}^0|K_1\rangle}\int\frac{d^4 k}{(2\pi)^4}\frac{1}{k^2-2p\cdot k}\nonumber \\
& &\times\{\langle K_1| \int d^4 x K_1^{\mu 0}\bar{K}_{1\mu}^0|K_1\rangle[4 m_{K_1}^2+(b^2+2b\gamma^2)k^2
+2\gamma^4 p\cdot k-\frac{4(p\cdot k)^2}{k^2}\nonumber \\
& &-\frac{1}{m_{K_1}^2}(b k^2-(b-\gamma^2)p\cdot k)^2]
+\langle K_1| \int d^4 x K_{1\mu}^0\bar{K}_{1\nu}^0|K_1\rangle k^\mu k^\nu [
-(3b^2-4 b+4)\nonumber \\
& &+D(b+\gamma^2)^2+4\gamma^2
-6b\gamma^2-2\gamma^4-\frac{2\gamma^4 p\cdot k}{k^2}
+\frac{1}{m_{K_1}^2 k^2}(b k^2-2(1-\gamma^2)p\cdot k)^2]\}\nonumber \\
& &\times k^2[\frac{1}{k^2-m_\rho^2}-\frac{1}{3}\frac{1}{k^2-m_\omega^2}
-\frac{2}{3}\frac{1}{k^2-m_\phi^2}]^2,\\
\vspace{2mm}
&&(m_{K_1^0}^2)_{EM}(3)
=\frac{-ie^2}{4\langle K_1| \int d^4 x K_{1\mu}^0
\bar{K}_1^{\mu 0}|K_1\rangle}\int\frac{d^4 k}{(2\pi)^4}\frac{1}{(p-k)^2-m_K^2}\nonumber\\
& &\times\{\langle K_1| \int d^4 x K_{1\mu}^0\bar{K}_1^{\mu 0}|K_1\rangle(\beta_1^\prime-3\beta_2 p\cdot k+
\beta_3 k^2)^2+\nonumber \\
& &\langle K_1| \int d^4 x K_{1\mu}^0\bar{K}_{1\nu}^0|K_1\rangle k^\mu k^\nu
[\beta_2 m_{K_1}^2-\frac{(\beta_1^\prime-2 \beta_2 p\cdot k+\beta_3 k^2)^2}{k^2}
]\}\nonumber \\
& & \times k^2[\frac{1}{k^2-m_\rho^2}-\frac{1}{3}\frac{1}{k^2-m_\omega^2}-
\frac{2}{3}\frac{1}{k^2-m_\phi^2}]^2.
\end{eqnarray}
where $p$ is the external momentum of $a_1$ or $K_1$-fields($p^2$=$m_a^2$ or
$m_{K_1}^2$ on mass-shell),
and
\begin{eqnarray*}
& &b=1-\frac{\gamma^2}{\pi^2 g^2},\\
& &\beta_1^\prime=\beta_1+(\beta_3+\beta_4-\beta_5) m_{K_1}^2.
\end{eqnarray*}
Taking $m_\rho=m_\omega=m_\phi$, and $m_a=m_{K_1}$, we obtain
\begin{eqnarray*}
(m_{a^\pm}^2)_{EM}(i)=(m_{K_1^\pm}^2)_{EM}(i),\;\;\;
(m_{K_1^0})_{EM}(i)=0,\;\;\; i=1,2,3.
\end{eqnarray*}
Then Eq.(35) or (2) holds, which indicates that Dashen's theorem can be generalized
to SU(3) sector of axial-vector mesons in the present theory.

Similarly, according to Refs.\cite{Li1,Li2}, the Lagrangians which contibute to
the electromagnetic masses of $\rho^\pm$ and $K^{*\pm}$ read
\begin{eqnarray}
{\cal L}_{\rho\rho\rho\rho}&=&-\frac{4}{g^2}\rho_\mu^0\rho^{0\mu} \rho_\nu^+
\rho^{-\nu}+\frac{2}{g^2}\rho_\mu^0\rho^0_\nu(\rho^{+\mu}\rho^{-\nu}+
\rho^{-\mu}\rho^{+\nu}),\\
{\cal L}_{\rho\rho\rho}&=&\frac{2i}{g}\partial_\nu \rho_\mu^0\rho^{+\mu}
\rho^{-\nu}-\frac{2i}{g}\rho_\nu^0 \rho_\mu^+(\partial^\nu \rho^{-\mu}-
\partial^\mu \rho^{-\nu})+h.c.,\\
{\cal L}_{\rho\omega\pi}&=&-\frac{3}{\pi^2 g^2 f_\pi}
\epsilon^{\mu\nu\alpha\beta}\partial_\mu\omega_\nu\rho_\alpha^+
\partial_\beta\pi^-+h.c.
\end{eqnarray}
\begin{eqnarray}
{\cal L}_{K^{*+}K^{*-}v v}&=&-\frac{1}{g^2}(\rho_\mu^0+v_\mu^8)^2 K^{+}_\nu
K^{-\nu}\nonumber \\
& &+\frac{1}{2 g^2}(\rho_\mu^0+v_\mu^8)(\rho_\nu^0+v_\nu^8)(K^{+\mu}K^{-\nu}+
K^{-\mu}K^{+\nu}),\\
{\cal L}_{K^{*+}K^{*-} v}&=&\frac{i}{g}(\partial_\nu\rho_\mu^0
+\partial_\nu v_\mu^8) K^{+\mu} K^{-\nu}\nonumber \\
& &-\frac{i}{g}(\rho_\nu^0+v_\nu^8) K^+_{\mu} (\partial_\nu K^{-\mu}-
\partial^\mu K^{-\nu})+h.c.,\\
{\cal L}_{K^{*\pm}K^\pm v}&=&-\frac{3}{\pi^2 g^2 f_\pi}
\epsilon^{\mu\nu\alpha\beta}K^{+}_\mu\partial_\beta K^-
(\frac{1}{2}\partial_\nu\rho_\alpha^0+
\frac{1}{2}\partial_\nu \omega_\alpha+\frac{\sqrt{2}}{2}
\partial_\nu\phi_\alpha)+h.c.
\end{eqnarray}
Eq.(48) and Eq.(51) come from the abnormal part of the effective Lagrangian $
{\cal L}_{IM}$(to see Refs.\cite{Li1,Li2}). Thus, similar to the above, it is not difficult to conclude that $
\rho^\pm$ and $K^{*\pm}$ can receive the same electromagnetic self-energies
in the chiral SU(3) limit,
\begin{equation}
({m_{\rho^\pm}^2})_{EM}=({m_{K^{*\pm}}^2})_{EM} 
\end{equation}

However, $\rho ^0$ and $K^{*0}$ can also obtain electromagnetic masses even
in the chiral SU(3) limit, which is different from the case of neutral
pseudoscalar and axial-vector mesons. The Lagrangian contributing to the
electromagnetic masses of $K^{*0}$ is 
\begin{eqnarray}
& &{\cal L}_{K^{*0} \bar{K}^{*0} v v}={\cal L}_{K^{*+}K^{*-} v v}
\{\rho^0\leftrightarrow -\rho^0, K^{*\pm}\leftrightarrow K^{*0}(
\bar{K}^{*0})\},\\
& &{\cal L}_{K^{*0} \bar{K}^{*0}  v}={\cal L}_{K^{*+}K^{*-}  v}
\{\rho^0\leftrightarrow -\rho^0, K^{*\pm}\leftrightarrow K^{*0}(
\bar{K}^{*0})\},\\
& &{\cal L}_{K^{*0} K^0  v}={\cal L}_{K^{*\pm}K^\pm v}
\{\rho^0\leftrightarrow -\rho^0, K^{*\pm}\leftrightarrow K^{*0}(
\bar{K}^{*0}), K^\pm\leftrightarrow K^0(\bar{K}^0)\}.
\end{eqnarray}
Note that in Eq.(55), the combination of the neutral vector mesons is $-\rho
_\mu +\omega _\mu +\sqrt{2}\phi _\mu $ instead of $-\rho _\mu +\omega _\mu -\sqrt{2
}\phi _\mu $ emerging in Eqs.(53)(54) and the Lagrangians contributing to
the electromagnetic masses of $K^0$ and $K_1^0$. Therefore, even in the
chiral SU(3) limit, the electromagnetic masses of $K^{*0}$ is nonzero due to
the contribution coming from Eq.(55)(the contributions of Eqs.(53) and (54) vanish in
the limit of $m_\rho =m_\omega =m_\phi $).

As to electromagnetic masses of $\rho^0$-mesons, the things are more
complicated. The contribution to $(m_{\rho^0}^2)_{EM}$ from ${\cal L}_{IM}$
is 
\begin{equation}
{\cal L}_{\rho\omega\pi}=-\frac{3}{\pi^2 g^2 f_\pi} \epsilon^{\mu\nu\alpha
\beta}\partial_\mu\omega_\nu\rho_\alpha^0 \partial_\beta\pi^0 
\end{equation}
Distinguishing from the case of $K^{*0}$, the direct $\rho^0$-photon
coupling which comes from VMD(Eq.(4)) can bring both the tree and one-loop
diagrams contributing to the electromagnetic masses of $\rho^0$ in the
chiral limit. Thus, $(m_{\rho^0}^2)_{EM}$ is nonzero. Furthermore, from the
point of view of large-$N_c$ expansion\cite{GH}, the tree diagrams are $O(N_C)$
while the one-loop diagrams are $O(1)$\cite{Li1}, in general, we can not
expect that $(m_{\rho^0}^2)_{EM}= (m_{K^{*0}}^2)_{EM}$ in the chiral SU(3)
limit(only loop diagrams can contribute to $(m_{K^{*0}}^2)_{EM}$).
So the generalization of Dashen's theorem fails for the vector mesons.
This result is in correspondence with one given by Bijnens and
Gosdzinsky\cite{JG}.

\vspace{8mm}

4. In summary, employing $U(3)_L\times U(3)_R$ chiral theory of mesons, we
obtain(in the chiral SU(3) limit) 
\begin{eqnarray*}
(m_{\pi^\pm}^2)_{EM}=(m_{K^\pm}^2)_{EM},\;\;\;
(m_{a^\pm}^2)_{EM}=(m_{K_1^\pm}^2)_{EM},\;\;\;
(m_{\rho^\pm}^2)_{EM}=(m_{K^{*\pm}}^2)_{EM}.
\end{eqnarray*}
As pointed out in Ref.\cite{Da}, because $\pi^{\pm }$ and $K^{\pm }$(or $a_1^{\pm
}$ and $K_1^{\pm }$, $\rho^{\pm }$ and $K^{*\pm }$) belong to the same
U-spin multiplet of the SU(3) group, their electromagnetic self-energies
must be equal in the chiral SU(3) limit. The electromagnetic masses of $\pi
^0$, $K^0$, $a_1^0$ and $K_1^0$ vanish in the chiral SU(3) limit. Therefore,
Dashen's theorem(Eqs.(1) and (2)) holds for both pseudoscalar and
axial-vector mesons. However, the contributions from the abnormal part of
the effective Lagrangian result in the nonzero electromagnetic masses of $
\rho ^0$ and $K^{*0}$ even in the chiral SU(3) limit, and VMD produces the
direct coupling of $\rho ^0$ and photon(Eq.(4)), which provides the another
contributions to electromagnetic masses of $\rho^0$. Generally, $(m_{\rho
^0}^2)_{EM}\not =(m_{K^{*0}}^2)_{EM}$. Hence, the generalization of Dashen's
theorem fails for vector-mesons.

Dashen's theorem is valid only in the chiral SU(3) limit. The violation of
this theorem at the leading order in quark mass expansion has been investigated
extensively\cite{DHW,JB,KK,RU,GYL}, and a large violation has been revealed in
Refs.\cite{DHW,JB,GYL}. In particular, a rather large violation of Eq.(2) has
been obtained in Ref.\cite{GYL}.

\begin{center}
{\bf ACKNOWLEDGMENTS}
\end{center}
D.N.Gao and M.L.Yan are partially supported by NSF of China through C.N.Yang.\\
B.A.Li is partially supported by DOE Grant No. DE-91ER75661.

\end{document}